\pdfoutput=1
\documentclass[acmlarge,nonacm]{acmart}
\usepackage{parskip}
\begin{document}
\title{Safety Co-Option and Compromised National Security: The Self-Fulfilling Prophecy of Weakened AI Risk Thresholds}
\author{Heidy Khlaaf}
\email{heidy@ainowinstitute.org}
\affiliation{%
  \institution{AI Now Institute}
}
\author{Sarah Myers West}
\email{sarah@ainowinstitute.org}
\affiliation{%
  \institution{AI Now Institute}
}

\authorsaddresses{}
\renewcommand{\shortauthors}{Khlaaf and Myers West}

\begin{abstract}
Risk thresholds provide a measure of the level of risk exposure that a society or individual is willing to withstand, ultimately shaping how we determine the safety of technological systems. Against the backdrop of the Cold War, the first risk analyses, such as those devised for nuclear systems, cemented societally accepted risk thresholds against which safety-critical and defense systems are now evaluated. But today, the appropriate risk tolerances for AI systems have yet to be agreed on by global governing efforts, despite the need for democratic deliberation regarding the acceptable levels of harm to human life. Absent such AI risk thresholds, AI technologists—primarily industry labs, as well as “AI safety”-focused organizations—have instead advocated for risk tolerances skewed by a purported AI arms race and speculative “existential” risks, taking over the arbitration of risk determinations with life-or-death consequences, subverting democratic processes.

In this paper, we demonstrate how such approaches have allowed AI technologists to engage in “safety revisionism,” substituting traditional safety methods and terminology with ill-defined alternatives that vie for the accelerated adoption of military AI uses at the cost of lowered safety and security thresholds. We explore how the current trajectory for AI risk determination and evaluation for foundation model use within national security is poised for a race to the bottom, to the detriment of the US’s national security interests. Safety-critical and defense systems must comply with assurance frameworks that are aligned with established risk thresholds, and foundation models are no exception. As such, development of evaluation frameworks for AI-based military systems must preserve the safety and security of US critical and defense infrastructure, and remain in alignment with international humanitarian law.

\end{abstract}
\maketitle

\section{Introduction}
At the height of the Cold War and on the heels of the Manhattan Project, US nuclear researchers devised the first risk-analysis frameworks to evaluate the risk and safety mitigations of developing and deploying powerful technological systems, while ensuring the US maintained an edge amidst a nuclear arms race with the Soviet Union. These established risk analyses and thresholds provided the precedent against which defense and safety-critical systems continue to be evaluated today~\cite{starr_philosophical_1976,starr_risks_1982,okrent_safety_1987}. Yet parallels of a purported AI-arms race with China are nevertheless being used to justify the jettisoning of these time-tested frameworks, at the cost of both safety and security, in exchange for the unfettered development and deployment of AI technologies~\cite{drexel_behind_2023,amodei_trump_2025,murgia_anthropics_2024}.

Establishing the accepted risk and dependability of AI systems has surfaced several challenges in the determination of how and if AI technologies can and need to be trusted in the context of safety-critical systems and accelerated military deployment. These challenges have often stemmed from a surfeit of hype, a lack of clear definitions, and seeming inability to clearly specify the precise intended application of AI systems. Global governance efforts, like those at the United Nations (UN), currently lack specificity, as the appropriate societal risk tolerances for AI systems have yet to be agreed on by civil society, legislatures, or other rulemaking and enforcement bodies. The absence of such concrete deliberation has given way for AI technologists to seize the determinations of said risk tolerances, and thus thresholds, going against long-established norms.

Indeed, despite invoking comparisons that AI risk is on par with safety risks that were once posed with the coming of the nuclear age~\cite{khlaaf_time}, and that have arguably resulted in the resilience of US military capabilities and critical infrastructure, \textbf{AI technologists are instead using the pretense of an AI arms race and speculative “existential” risks to discard the very risk and safety thresholds established by this era}. The purported, and even dispelled, cost-benefit justification for this discrepancy is that the accelerated and ubiquitous adoption of AI above all else provides a marker of the US’s technological advantage and defense prowess over China and other adversaries~\cite{amodei_trump_2025,drexel_behind_2023,hendrycks_nuclear-level_2025,murgia_anthropics_2024}. Not only are these presumptions unsubstantiated given the historical inaccuracies and questionable efficacy of AI-based systems~\cite{boulanin_risks_2025,cummings_ai_2020,khlaaf_mind_2024,viveros_alvarez_risks_2024}, but they also fully concede the determinations of risk tolerances and thresholds to AI technologists, reflecting an adoption of risk scales that are fundamentally at odds with democratic deliberative norms. Put bluntly, the militaristic push for AI adoption puts life-or-death decisions in the hands of technologists instead of governing and legislative bodies, often in ways that are opaque and unaccountable, while defying previous precedents established for novel technological advancements. These matters extend beyond influencing how AI is used on the battlefield: Military safety-engineering standards\footnote{These are standards adopted by the military that define appropriate practices for guaranteeing system safety. For example, DOD MIL-STD-882 outlines the US DOD’s approach to “eliminating hazards, where possible, and minimizing risks where those hazards cannot be eliminated,” and defines thresholds for acceptable risk~\cite{noauthor_standard_2012}.} can set precedents for the evaluation of civilian safety-critical systems (and vice versa)~\cite{NATO_joint_2018}, thus shaping broader standards that define the infrastructures that members of the public interact with every day. 

Ironically, this approach—while articulated as necessary for national security—not only shifts institutional risk framing to bolster the embedding of technological systems with autocratic values within critical US infrastructure, but also, \textbf{by allowing for accelerated AI adoption at the cost of lowered safety and security thresholds, may be precisely what disadvantages US military and technological capabilities against China}. As such, a slew of existing national-security concerns emanating from uses of AI models will be ripe for adversaries to exploit and compromise US national and defense infrastructure. If integrated into critical infrastructures, AI’s persistent lack of performance may lead to strategic, operational, and tactical mistakes that could compromise the US’s international standing, and have deadly and geopolitically consequential impacts. Indeed, a plethora of critiques have explored the destabilizing impact of the AI arms race on geopolitics, international relations, and AI regulation~\cite{belfield_why_2022,cave_ai_2018,haner_artificial_2019,kak_2023_2023,roff_frame_2019,toner_illusion_2023,weiss_china_2022}. However, such critiques have not yet addressed how shifted risk thresholds diminish the role safety plays in the assurance of critical systems by weakening (purposely or inadvertently) long-established safety-engineering principles that would deem many AI systems unsafe and insecure for military use. As such, discarding rudimentary safety-engineering measures to evaluate AI-enabled military systems will imperil not only the safety and security of military personnel, but of the civilian populace as well.

In this paper, we examine the history of safety engineering and its parallels with today’s AI arms race, highlighting the appropriation and co-option of safety engineering and verification and validation (V\&V) terminology within the field of “AI safety” to justify unverifiable and unsubstantiated claims regarding AI capabilities—claims that substitute technical and regulatory terms such as “safety,” “safety cases,” and “red-teaming” with substantially revised definitions that hollow out assurance methodologies, ultimately compromising safety-critical and defense systems. As an alternative, we offer a grounded exploration of AI safety that reconnects this term with its disciplinary origins. This work seeks to help remedy an alarming gap in the field by bringing military AI systems to the forefront of the AI safety and security discussion. In particular, if evaluation of foundation models serve as the basis of prospective assessments with which emerging AI safety-critical and defense systems are to comply (i.e., lethal autonomous weapons systems[LAWs] or AI decision support systems[DSS]), then as they stand they must be aligned with and assured against risk thresholds that continue to maintain the safety and security of US critical and defense infrastructure, while remaining in line with international humanitarian law (IHL).

\section{Historical Determinations of Risk Thresholds and AI}

The safety\footnote{We use the term “safety” in the traditional systems-engineering sense, meaning the prevention of a system from impacting its environment in an undesirable or hazardous way, typically aiming to protect human lives, the environment, or even military assets. This is distinct from the meaning of the term as used in AI, which we will address in more detail in the following sections.} of technological systems is a well-established discipline spanning several fields that is often revisited with the advent of a novel technology, necessitating the adaptation of existing methodologies for newly introduced hazards. Establishing the safety of AI systems, and in particular foundation models, has reinvigorated a focus on safety methodologies such as risk assessments and safety cases in pursuit of the assurance of AI systems given a surge of interest in deploying them widely, including within national infrastructure~\cite{noauthor_artificial_nodate}. Yet despite numerous efforts to adopt safety practices to address the unique hazards introduced by AI systems~\cite{anderljung_frontier_2023,bloomfield_assurance_2024,khlaaf_hazard_2022,khlaaf_toward_2023,raji_concrete_2023, habli_big_2025,deepmind_introducing_2025}, these works have failed to address a key gap in the AI safety discussion: establishing the risk thresholds of AI systems that safety assurances are to be determined against. That is, adequately addressing \textbf{the question of “How safe is safe enough?” first requires establishing a societally accepted risk-benefit trade-off or tolerance that provides safety goals and thresholds against which technological systems can be assessed, a discussion largely absent from current AI governing efforts.} 

In this section, we surface historical lessons on risk tolerances and thresholds established for technologies such as nuclear energy that provided a precedent against which modern-day safety-critical systems continue to be evaluated. We then examine how, in failing to determine safety goals that follow tried-and-true approaches, AI global governance efforts have allowed AI promoters to instead supplant risk tolerances with an AI-arms-race logic that ultimately allows for accelerated AI adoption at the cost of lowered safety thresholds. In the ensuing sections, we demonstrate how the absence of societally accepted risk tolerance for AI has also allowed AI technologists to wield safety and V\&V terminology toward unverifiable or speculative risk thresholds and claims that intend to signal compliance with established safety and defense thresholds, whilst only displaying a veneer of safety.

To effectively demonstrate the gap in which AI safety frameworks fail to adequately address “How safe is safe enough?” and the safety goals needed to establish the acceptable risk to the public, we first revisit the foundational work of Chauncey Starr, a nuclear and electrical engineer and one of the pioneers of probabilistic risk analysis. Through his work on the Manhattan Project and subsequent efforts with Atomics International to pursue the commercialization of the generation of electricity by nuclear power, Starr formalized the first risk analysis frameworks to evaluate the risk and safety mitigations of developing and deploying powerful nuclear technological systems amidst a nuclear arms race with the Soviet Union~\cite{starr_risk_1981,grant_chauncey_2007,nrc_safety_1983,victor_k._breeder_1973}. Working under geopolitical conditions that mirror our own, Starr was especially privy to the ongoing nuclear arms race at the time, and even believed that President Truman was correct in his decision to use atomic weapons against Japan; Starr himself had helped develop the “Little Boy” bomb~\cite{power_chauncey_2007}. \textbf{Despite this, Starr’s formulation of risk thresholds and analysis significantly deviates from the arms race calculus being touted today by AI technologists.}

Specifically, Starr defined risk analysis to be the study of the relevant cause-and-effect relationships that give rise to safety risks, to describe their magnitudes and distributions, and to identify mitigations for risk reduction, which concerns the character of risk and the social significance of risk identified~\cite{starr_philosophical_1976}. He posited that risk analysis would require the consideration of the societal evaluation of risk and, as such, interpreting public attitudes and values. This characterization was intended to emphasize that risk analysis concerns at once questions that can be addressed scientifically and questions involving public policy concerns (i.e., societal safety)\footnote{We note that modern-day uses of the term “risk assessment” may not address societal decisions carried out on systems for which risk thresholds have already been determined.}—where \textbf{risk tolerance} characterizes the degree, amount, or volume of risk that a society or individual will withstand, while \textbf{risk threshold} measures the level of risk exposure above which action must be taken to address risks proactively, and below which risks may be accepted. Starr noted that risk acceptance, and thus thresholds, are inseparable from risk perception and evaluation. 

Starr observed two primary, and often conflicting, societal value systems of risk that reflect different types of social costs to risks: \textbf{societal risk} and \textbf{individual risk}. In modern-day literature, societal risk has been formalized to concern the probability of an event with many fatalities; individual risk has been formalized to concern the annual probability of death of a single person. Starr provided a more intuitive and example-driven distinction, where an individual value system “charges an increasing social cost to each successive disability day.” Under such a system, the social cost of one individual being disabled or harmed for a year is considered greater than the cost of 365 separate individuals each disabled or harmed for one day.

With regard to selecting either societal or individual value systems of risk, a determination of risk tolerance and thresholds is intended to be driven by societal decision-makers that bridge the gap between the perceived and assessed risks and benefits of a novel technological system. Starr pointedly noted that autocratic powers and military organizations overindex societal risk thresholds in the aim of preserving the welfare of these regimes rather than an individual risk value system, whereas democratic governments use a mix of both scales to determine thresholds of safety. These dynamics extend to the role of technologists themselves in the establishment of risk thresholds: Starr noted that in a democratic society, a technologist’s role ends with providing the quantification of risk impacts, as has often been reflected by US advisory mechanisms that have allowed science advisers to shape, with limitations, US federal policy with key technologies such as nuclear~\cite{jasanoff_fifth,mackenzie_inventing_1990}. Otherwise, Starr characterized the placement of technocratic experts in decision-making positions that extend their role beyond risk assessors as indicators of an autocratic society~\cite{starr_philosophical_1976}.

Indeed, the historical determinations of safety thresholds for safety-critical systems within the US have largely been established through broad democratic deliberations involving elected representatives, regulators, and expert consensus-building that draws on both societal and individual value risks. This has included risk management to implement policies consistent with values as perceived by democratic institutions, and managing public risk perceptions through mitigations, including for nuclear systems even against the backdrop of the Cold War~\cite{jasanoff_containing_2009}.\footnote{An example of such threshold includes Starr's recommendation for nuclear power where “the upper bound be set at a risk level equivalent to those risks of routine living which are normally accepted, i.e., about 10\textsuperscript{-4} deaths per year per person (100 deaths/yr/million). The proposed lower design target is 10\textsuperscript{-8} (0.1 deaths/yr/million), about one-hundredth of the minimal risk from the natural hazards all people are exposed to”~\cite{starr_risk_1981}.}

\textbf{We find ourselves in dramatically different circumstances with today’s AI development and deployment efforts}. Rather than through democratic deliberation, AI technologists are exclusively seeking to establish the risk tolerances—and thus thresholds—by which AI systems are evaluated, utilizing skewed risk and safety interpretations at odds with the demands of social, political, and ethical civilian life~\cite{schwarz_silicon_2021}, and with assurance practices established by experts like Starr that are critical to the safety and security posture of national and defense infrastructure. Under the pretense of an AI arms race and “existential” risks, “AI safety” proponents are working to accelerate the adoption of AI within defense and critical infrastructure at the cost of safeguarding implementation—justifying likely harm to both defense personnel and civilians by the need to presumably establish the US’s technological advantage and defense prowess over China~\cite{amodei_trump_2025,drexel_behind_2023,hendrycks_nuclear-level_2025,murgia_anthropics_2024}. Yet the parallels between nuclear proliferation and AI development have been found to be misrepresented and largely unfounded, not only because of the lack of definable success of AI within military applications that have relied on perception rather than fit-for-purpose capabilities, but also because AI is not itself a weapon; it is, rather, a decision support or automation function~\cite{cummings_ai_2020,roff_frame_2019}. Unsubstantiated claims aside, the prioritization of militarism in current AI policymaking in the US indicates a shift toward a societal value for risk determination rather than one of individual value, reflecting an adoption of risk scales associated with autocratic rather than democratic societies that would require democratic deliberation on the impacts of AI and acceptable thresholds for risks.

In such an autocratic society, Starr noted, the social costs of harms or accidents are determined (whether accurately or not) to be directly proportional to the estimated expected values of direct harm costs, and as such “a social value system relating to risk characteristics is not applied, and the decision process is rather automatic”~\cite{starr_philosophical_1976}. We see this reflected in the recent flurry of AI-driven systems being deployed to purportedly automate bureaucracy and critical and defense governing decision-making ~\cite{noauthor_canada_nodate,noauthor_britain_2025,noauthor_artificial_nodate,mason_ai_2025,wong_doges_2025}. Such processes implicitly position risk assessors (i.e., AI technologists) as decision-makers by eliminating the channels through which public attitudes and values can bridge the gap between the perceived and assessed risks and benefits of novel technological systems, laundering decision-making and risk determination to the very AI systems built by technologists. For example, lofty claims about potential AI use are being deployed to displace grounded expertise, such as in the case of the US Department of Government Efficiency (DOGE), where the promise that AI can administer large parts of the federal government has functioned to undermine and discredit the expertise of career civil servants and to present this as a technical transition, rather than as a political transfer of power~\cite{chen_dispelling_nodate,salvaggio_anatomy_2025,kelly_elon_nodate}. This is further bolstered by the lobbying power of AI industry players who influence decision-makers within the US in ways that go beyond the limitations typically placed on US science advisers in shaping federal policy~\cite{bordelon_how_2023,bordelon_key_2023}. \textbf{Starr’s frameworks thus dispel claims that the acceleration of AI use within the national security apparatus would allow “democracies to maintain the lead”: The means themselves are reflective of a transformation away from democratic use of technology and toward the very autocratic standards that these technologists warn against in a purported AI-arms race with China.}

Finally, the value judgments required to convert risk decision factors to a common AI risk threshold metric have been muddled by a surfeit of hype, a lack of clear definitions, and a concomitant absence of clearly specified claims about the precise intended application of AI systems. In neglecting to consider that safety frameworks can only be effective when operating within the parameters of a societally accepted risk value system, research and global governance efforts alike have allowed diffuse conversations regarding “AI safety” to substitute the concrete and precise language used in safety engineering with interpretations antithetical to their original meaning.

We note that various risk-perception literature, including Starr’s, does not directly compare how technological system risks compare to human errors, as the safety-critical systems and processes under scrutiny (e.g., financial, health, nuclear risks) are not ones that human labor could substitute, and thus have their risk be compared against (as in the case for some sociotechnical systems). However, recent AI developments have been explicitly concerned with the automation of human labor (e.g., foundation models), leading some to demand that standardized risk analyses be overhauled with measures only leveraging average human performance as the acceptable risk threshold. Yet safety-critical risks concern the lack of harm or failures arising in an attempt to meet a system intent, and not the intent itself. \textbf{Given that AI failure modes are distinctly and unpredictably different from that of humans’, their risk categories fall firmly within the scope of traditional risk frameworks aimed at safety-critical technological systems, regardless of whether or not an AI system is intended to automate human capabilities.} This issue has been explored in other capacities, including within the autonomous-driving literature, notably regarding flawed and misleading human performance metrics used to justify insufficient AI reliability~\cite{stilgoe_how_2021,cummings_identifying_2024,kalra_driving_2016}. Given the safety-critical context of AI use within defense, we thus exclude human performance as an appropriate or feasible risk metric for use in our analysis, and revisit the impacts of such metrics on benchmarks in later sections.

In the next section, we demonstrate how these skewed risk tolerances have allowed for the co-option of safety methodologies through permitting the use of unverifiable claims, or claims that are not indicative of the accepted safety thresholds that have governed other technologies. We also explore how, in neglecting to apply safety thresholds that have been established since the nuclear arms race, global governance efforts have allowed for implicit acceptance of arbitrary risk tolerances and thresholds, as determined by those with self-interest in the development and deployment of AI-based systems. As a result, this safety revisionism will not only undermine and fail to preserve national security and its necessary compliance with IHL, but will imperil civilian safety-critical systems to the detriment of the public.

\section{Safety Revisionism and Implications for National Security}

Within the AI discourse, the term “safety” has come to have a multitude of definitions that vary according to the context and the community. Yet these definitions seldom overlap with the broader meaning of “safety” used within safety-critical fields, including in defense, and may in fact directly contradict it. “Safety” in AI has become conflated with “alignment” or “capabilities”~\footnote{The use of the term “capabilities” in AI is not to be confused with its definition in the systems engineering context, which is the ability to execute a specified course of action given a set of requirements.} that aim to steer AI systems toward human-oriented intents and goals and prevent a “model’s capabilities from proliferating broadly”~\cite{anderljung_frontier_2023,deepmind_introducing_2025,noauthor_anthropics_2023,brown_value_2021,qi_fine-tuning_2023,clymer_safety_2024}. These metrics are not only subjective given their reliance on hypothetical scenarios untethered to empirical assessments, but fundamentally conflate safety properties with system intent~\cite{khlaaf_toward_2023}. Safety must center the protection of human lives, the environment, and critical assets. This is achieved by ensuring that the system is prevented from impacting its environment in an undesirable or hazardous way, including due to the system intent itself. Safety is also context-sensitive and there are no standard assurance or assessment approaches for “general” safety (as posited by AI literature), since doing so would render the evaluation of risk and safety intractable ~\cite{khlaaf_toward_2023, bloomfield_assurance_2024,mcdermid_upstream_2024}. Ultimately, the objective of safety is “freedom from risk which is not tolerable”[ISO/IEC Guide 51:1999, definition 3.1]; or, in other words, residual risk is reduced as far as reasonably tolerable or practicable.\footnote{This is a principle known as “as low as reasonably practicable” or ALARP.}

The discrepancy is not an issue of pedantry. Alignment and capabilities measures are simply not concerned with the system hazards arising out of specified intent, but rather with a system’s intent itself, either through the speculative ascription of autonomy to foundational models and their ability to inflict harm, or through concerns of how deployed models may be “broadly” misused. Concerns of speculative capabilities aside (given their lack of verifiability and thus applicability to safety frameworks), considerations of how novel technologies may be misused are not novel, whether they regard nuclear proliferation or use of software to commit cyber crimes~\cite{habli_ai_2024,narayanan_ai_nodate,doomed_crypto,jasanoff_containing_2009,orben_fixing_2025}. \textbf{Although adapting systematic safety thinking to capability evaluations can provide constructive outcomes ~\cite{khlaaf_hazard_2022,mcdermid_upstream_2024}, these are not context- or safety-specific hazards or failure modes as traditionally defined under safety-engineering frameworks. As such, their conflation with safety risks impacts the efficacy and relevance of security and safety engineering techniques necessary for the assurance of AI within mission-critical systems (e.g., defense)} Yet the supplanting of traditional safety assessments by broad “capabilities evaluation” frameworks is precisely what AI technologists and labs are proposing for evaluations in their pursuit to deploy foundation models within safety-critical and defense applications for use in DSS, intelligence, surveillance, target acquisition, and reconnaissance (ISTAR), or even within counter-unmanned aircraft systems~\cite{noauthor_anduril_nodate,noauthor_palantir_nodate,vincent_inside_2023,noauthor_thunderforge_nodate,noauthor_u.s._2024,khlaaf_mind_2024,primer_2023}. Indeed, the absence of societally accepted AI risk tolerances has allowed AI technologists to leverage both the altered and unfounded definition of safety, jointly with skewed AI-arms tolerances, to bolster the deployment of frameworks that deviate from well-established military standards necessary for the functionality of defense systems.

To demonstrate how this conflation undermines safety, consider that the traditional definition of safety was first conceived to ensure the performance of safety-critical infrastructure and defense systems~\cite{mcdermid_upstream_2024}, where dependability properties (e.g., functionality, performance, reliability, operability, robustness, availability, and even security-informed-safety~\cite{kaur_dependability_2018,noauthor_security-informed_2023}) have since become fundamental requirements for safety-critical applications. Despite this, recent rhetoric that “the AI future is not going to be won by hand-wringing about safety. It will be won by building, from reliable power plants to the manufacturing facilities that can produce the chips of the future" ~\cite{house_read:_2025,noauthor_britain_2025}, effectively demonstrate how the AI “alignment” and speculative “capabilities” narrative has distanced the term “safety” from its context, despite its historical centrality and necessity in assuring critical and defense systems. \textbf{This enables safety to be rendered as “hand-wringing,” working to the detriment of methodologies that critically ensure the functionality, reliability, operability, robustness, and availability properties of safety critical systems}—the same “reliable power plants” Vance calls for—and precisely demonstrating the harms in revisionism of safety and security terminology.

Deviations from traditional safety terms not only break with established safety thresholds and corresponding safety and dependability properties, but render safety techniques ineffective, as has also been demonstrated with the use of “safety cases” as redefined by AI technologists. Traditionally, a safety case is a structured argument supported by evidence intended to justify that a system is acceptably safe for a specific application in a defined operating environment. A safety case is often used as a regulatory mechanism for the basis of approvals ~\cite{kelly_systematic_2004}. Safety cases, and safety analyses more generally, are intended to model causal dependencies between faults and failures, and functional capabilities, in the case of safety of the intended functionality~\cite{mcdermid_upstream_2024}. In defense, the use of a safety case enables the understanding of the cumulative and interrelated risks associated with the use or deployment of a system~\cite{noauthor_implementation_nodate}. Yet AI labs have redefined this methodology with concepts such as “safety cases with red-team validation” and “affirmative cases.” The former is defined as “a robustness target set based on a safety case considering factors like the critical capability and deployment context,” while the latter is intended to support claims regarding “models not autonomously attempt[ing] to strategically undermine our safety measures or cause large-scale catastrophe”~\cite{deepmind_introducing_2025,noauthor_anthropics_2023}. Both definitions are at odds with traditional safety cases and risk thresholds.

For one, a safety case is not intended to produce any “targets” or thresholds, as defined by Google DeepMind’s “Frontier Safety Framework”~\cite{deepmind_introducing_2025}, but rather is the mechanism to substantiate that a system meets a defined claim, one that is typically derived from a risk threshold. As previously noted, risk thresholds are derived from risk tolerances that characterize the degree, amount, or volume of risk that a society or individual will withstand for a specific application. As such, these risk thresholds are established through risk analyses as initially formalized by Starr~\cite{starr_philosophical_1976,starr_risks_1982} and not through safety cases. In Table~\ref{tab:sil-table}, we provide an example of risk thresholds expressed in Safety Integrity Levels (SIL), a formal measure of expected system safety performance. These revised definitions thus blur the distinction between the processes through which risk thresholds are defined (i.e., risk analyses) and the safety analyses through which risk thresholds are substantiated (e.g., safety cases). This lack of distinction undermines not only the soundness of these methodologies, but also the independence between the democratic deliberation of risk tolerances and the technologists that provide the quantification of risk impacts, as noted by Starr~\cite{starr_philosophical_1976}. 

DeepMind’s framework also casts safety cases as a mitigation measure, when safety cases are in fact the primary activity that enables the exploration of how a system impacts its environment in an undesirable or hazardous way (e.g., hazard analysis), and for the demonstration of how a system implementation and hazard mitigations enable a system to meet an acceptably safe and predetermined risk threshold for a specific application. Other works have likely misinterpreted criticism of confirmation bias in safety cases with the inability to consider hazards and harms~\cite{noauthor_anthropics_2023,clymer_safety_2024}, yet this criticism itself has long been addressed through the use of “defeater” claims~\cite{defeaters}. Indeed, safety cases encompass all the necessary activities to substantiate or counter safety claims, and, as such, substitution of well-established safety activities within a case (e.g., through Hazard and Operability Studies, Fault Tree Analysis, Failure Modes and Effects Analysis) with ill-defined methodologies such as “red-teaming”~\cite{jones_ada,khlaaf_toward_2023}, will only undermine a safety case’s efficacy in determining that a critical system is acceptably safe.

\begin{table}[]
\begin{tabular}{llll}
Safety Integrity Level & Safety               & Probability of Failure on Demand & Risk Reduction Factor \\
SIL 4                  & \textgreater 99.99\% & 0.001\% - 0.01\%                 & 100,000 - 10,000      \\
SIL 3                  & 99.9\% - 99.99\%     & 0.01\% - 0.1\%                   & 10,000 - 1,000        \\
SIL 2                  & 99\% - 99.9\%        & 0.1\% - 1\%                      & 1,000 - 100           \\
SIL 1                  & 90\% - 99\%          & 1\% - 10\%                       & 100 - 10             
\end{tabular}
\caption{A SIL determines the risk-reduction factor required to ensure that the risk associated with hazardous events is reduced to acceptable quantitative levels through system-level or operational mitigations.}
\label{tab:sil-table}
\end{table} 

Finally, proposed “frontier safety frameworks” have supplanted traditional safety claims with assertions concerning “capabilities risks”~\cite{noauthor_anthropics_2023,deepmind_introducing_2025,irving_AISI} that fail to justify their safety targets and involve deliberative or consensus-building approaches to establish an accepted balance between societal and individual value risk scales.\footnote{Recent works have addressed similar shortcomings and have attempted to refine the expressiveness of “general” safety arguments to be more conducive to verifiable argumentation~\cite{habli_big_2025}. However, this work similarly lacks concrete risk thresholds that foundation models can be assured against, and problematically cedes the acceptability of said risk tolerances to “AI advisors.”} Indeed, a safety case must demonstrate that the safety targets are well-founded and that the safety functions implemented within a system meet said targets with consideration of the anticipated hazards associated with operating a system, neither of which are demonstrated within these proposed frameworks. Examples of such claims include AI models “not autonomously attempt[ing] to strategically undermine our safety measures” or “not caus[ing] large-scale catastrophe”~\cite{deepmind_introducing_2025,noauthor_anthropics_2023,clymer_safety_2024}, without further clarification on the quantitative thresholds they present~\cite{Kasirzadeh_2024} nor specificity on what “catastrophe,” “undermine,” or “autonomously” mean. These claims not only utilize the problematically revised definition of “safety,” but lack the formalization and specificity of risk-tolerance quantification that would be necessary to produce verifiable risk-reduction arguments (see Table~\ref{tab:sil-table}) and the system-level evidence to substantiate them beyond hypothetical scenarios. Although safety case notations such as Claims, Arguments, Evidence (CAE) can be utilized toward non-safety goals and as a tool for general reasoning (i.e., assurance cases), the use of the term “safety case” in the context of the assurance of AI for defense and national security signals an alignment with established military standards and terminology that proposed “frontier safety frameworks” simply do not satisfy. Indeed, safety in defense and safety-critical systems would necessitate context, and as such \textbf{the safety of AI systems cannot be effectively evaluated in the abstract, and assessments must address either specific use cases and contexts or deployment given a range of determined risk thresholds.}

But absent AI risk thresholds established through democratic deliberation, and military initiatives that place AI technologists in positions to form corresponding AI evaluation frameworks~\cite{noauthor_task_2024,noauthor_thunderforge_nodate,amodei_trump_2025,noauthor_anthropics_2023,announce_AISI}, AI technologists have sought to substitute traditional safety frameworks with the aforementioned ill-defined “capabilities” or “alignment” counterparts that deviate from well-established military standards necessary for the functionality of defense systems and ultimately cede the determination of risk tolerance to AI technologists. Put bluntly, this places decision-making regarding the acceptable risk tolerance for AI, including number of deaths, in the hands of AI technologists themselves (a signifier of an autocratic risk value system). Yet they have insufficiently shown the responsibility and rigor necessary for this task in the context of national security and where risk tolerance needs to be accountable to IHL.

\subsection{Impacts on National Security}
This safety revisionism has significant consequences for the defense and national security apparatus. Not only do revised definitions of “safety” and “safety cases” deviate from well-established military standards necessary for the functionality of defense systems, but the speculative nature of “AI safety” claims often hinges on unfounded assumptions regarding the capabilities and “intent” of AI systems that may be leveraged to justify the deployment of risky systems. For one, the misuse of terminology intended to signal compliance with established safety and security defense practices can mislead duty holders\footnote{A defense term for those responsible for the management of risk.} regarding the claims that an AI system complies, and can lead to approval of AI systems that display only a veneer of safety. Furthermore, unsubstantiated and even dispelled~\cite{toner_illusion_2023,cummings_ai_2020,roff_frame_2019,cave_ai_2018} claims of capabilities have already been used to advocate for a skewed cost-benefit calculus, where the accelerated adoption of AI above all else, including the dependability of these systems, has been touted as a marker for the US’s technological advantage and defense prowess over China and other adversaries~
\cite{amodei_trump_2025,drexel_behind_2023,hendrycks_nuclear-level_2025,murgia_anthropics_2024,noauthor_palantir_nodate}. Weakened terminology and arms-race justification come at a clear cost, as identified risks and failure modes continue to be unveiled~\cite{weidinger_ethical_2021,carlini_stealing_2024,nasr_scalable_2023,carlini_poisoning_2024,zou_universal_2023,zhang_persistent_2024,carlini_extracting_2023} despite purported claims that “potential” AI systems will eventually eliminate these risks. 

Rhetoric that has attempted to prioritize security threats over safety hazards will only escalate national-security failures, as security and safety are closely related and their interactions introduce complex consequences of failure and interdependencies~\cite{noauthor_security-informed_2023}. Examples of such risks include safety hazards emanating from the inability to prevent personally identifiable information from contributing to the use and proliferation of defense AI capabilities such as ISTAR by adversaries~\cite{khlaaf_mind_2024}. They also include interconnected security hazards, where it has been demonstrated that the usage of foundation models within military settings inherently expands the attack vectors military systems are susceptible to, including the defense infrastructures they interface with, particularly given new AI supply-chain vulnerabilities. These supply-chain threats can lead to novel attack surfaces that allow unauthorized access and changes in data, model parameters, and intellectual property given a lack of separation between commercial models and those deployed within military applications~\cite{khlaaf_mind_2024,carlini_stealing_2024,nasr_scalable_2023,carlini_poisoning_2024,zou_universal_2023,zhang_persistent_2024,carlini_extracting_2023}. These vulnerabilities pose defense and national security risks due to the significant increase of attack surfaces that AI-based systems present, which can inadvertently or maliciously accelerate these systems toward a state of imprecise and unlawful targeting and a loss of human control, impacting both the safety and security of defense infrastructure and also creating conditions that may harm defense personnel and give rise to potential violations to IHL~\cite{icrc_2024}.

If appropriate military standards and safety procedures had been applied in pilots that have sought to integrate foundation AI models in defense use, then these risks would be easily unveiled. This has not been the approach taken by the US Department Of Defense (DOD) Task Force Lima, which was established to create an evaluative framework for the deployment of foundation models in defense contexts. Rather than create a framework stemming from traditional safety and defense practices, Task Force Lima took on a “capabilities evaluation” approach that was then concluded through the creation of an Artificial Intelligence Rapid Capabilities Cell (AI RCC) designed for “rapid user-centric experimentation”—though details on the methodologies that will be used for system testing are slim. Similar approaches have been taken by other governments, such as the UK’s “AI Security Institute”~\cite{irving_AISI,announce_AISI}, despite important assertions of caution by leaders in the defense community; for example, the Navy’s chief information officer, Jane Overslaugh Rathbun, noted that commercial models have “inherent security vulnerabilities” and are “not recommended for operational use cases”~\cite{navy_department}. In the next section, we further address the chasm between “capabilities evaluations” of foundation AI models and their inability to meaningfully demonstrate their safety, security, and fitness for purpose by examining inconsequential evaluation criteria that have been scoped too narrowly or have little to no relevance for defense applications, and ultimately break with defense V\&V norms.
\section{The Facade of Capabilities via Manufactured Metrics}

The principles of evaluation for defense and safety-critical applications hinge on the well-established formalization of Test \& Evaluation, Validation \& Verification (TEVV), where verification ensures that the system meets safety specifications as designed, while validation confirms that the system fulfills its intended purpose in its operational environment without causing unacceptable risk~\cite{noauthor_dod_2024}. That is, TEVV processes are intended to evaluate the functional requirements of a system, including its reliability, operability, robustness, and availability, while also ensuring that a system meets its safety requirements intended to avert risk. However, evaluation and benchmarking criteria for foundation AI models have significantly deviated from these tried-and-true approaches in favor of ill-defined methodologies such as “red-teaming” and general evaluations~\cite{noauthor_task_2024,noauthor_u.s._2024} that are scoped too narrowly or have little to no relevance for safety-critical applications~\cite{khlaaf_toward_2023,jones_ada}, whilst continuing to be propped up by a flawed arms-race calculus. Such an approach will ultimately compromise strategic, operational, and tactical defense missions given their proposed use to support system-specific claims that they cannot substantiate. 

Once again, the absence of enforceable risk thresholds has led to an implicit acceptance of arbitrary risk tolerances and assessments as determined by those with self-interest in the development and deployment of AI-based systems, without independent scrutiny. The combination of corporate secrecy shielding AI developers with military classification shielding deployed use exacerbates this problem. The trend toward the use of “general” benchmarking for defense applications, wherein a set of arbitrary tasks is intended to indicate the generalizable capabilities and risks of an AI model, demonstrates a profound misunderstanding of the foundations of TEVV, specifically that safety-critical and defense systems cannot be substantiated without application-specific requirements~\cite{khlaaf_toward_2023,cisa_ai_2024,bloomfield_assurance_2024,mcdermid_upstream_2024,noauthor_standard_2012}. 

Consider type certification in aviation, where evaluations are carried out for the approval of a particular vehicle design under specific airworthiness requirements (e.g., Federal Aviation Administration 14 CFR part 21). There is no standard assessment approach for “generic” vehicle types across all domains given varying requirements and accepted risk threshold depending on the deployment environment of a vehicle. It would thus be contrary to established practices—and unproductive—to presume that evaluating foundation models against puzzle-solving tasks or a subset of generic tasks would be generalizable to the performance of systems designed for mission-critical tasks such as command and control (C2), weapons development, intelligence gathering, target acquisition, and reconnaissance. For the evaluation of hazards, “red-teaming” in AI has come to refer to probing foundational models for hazardous or harmful outputs, which are then used to update the model safeguards latently~\cite{ganguli_red_2022,ouyang_training_2022,perez_red_2022}. Note that “red-teaming” is not defined according to its traditional security context.\footnote{In cybersecurity, the intent of a red-teaming exercise is to realistically test an organization’s capability to detect and respond to a staged adversarial attack, and to assess and validate security posture and attack resilience.} The equivalent technique under TEVV that was likely intended for use is boundary- or stress-testing, a verification technique that aims to test edge cases or fringe inputs that may lead to unknown failure modes and potential hazards. However, just as with the revisionism of “safety” and “safety cases,” AI “red-teaming” fails to consider the context against which potential harms and hazards are to be measured, and as such cannot substantiate safety and security requirements intended to avert risk as would be required by TEVV. Furthermore, the safeguards that result from red-teaming exercises have been consistently demonstrated to be insufficient and brittle, bringing into question their applicability in safety-critical environments~\cite{zou_universal_2023,zhang_persistent_2024,jones_ada}.

The correct selection and use of TEVV methodologies is paramount to the assessment of novel military technologies in assuring their fitness for purpose to prevent casualties whilst maximizing the destruction of intended adversaries within the bounds of IHL. Yet novel military technologies alone, from the “electronic battlefields” in Vietnam to drone warfare in the Middle East, have never been able to replace a realistic assessment of what military force can achieve~\cite{quincy_private_2024}. Today, similar challenges arise with proposed AI-based defense technologies that not only remain untested, but are highly prone to failures amid purported and contradictory claims that AI-driven systems minimize civilian casualties~\cite{abraham_lavender:_2024, biesecker_as_2025,dwoskin_israel_2024,quincy_private_2024}. As previously noted, AI failure modes are distinctly and unpredictably different from human failure modes and, as such, the assessment for their fitness would not only require affirmation of their functionality (i.e., validation), but also assurance that AI models meet safety requirements intended to avert risk that may arise due to their failures (i.e., verification). Indeed, AI systems pose an intractable number of risks relative to previous technological counterparts given their nondeterministic nature, and the trend to train AI models for general tasks (i.e., foundation models) rather than military-exclusive functionality only expands the failure modes that must be accommodated for further~\cite{khlaaf_mind_2024,khlaaf_toward_2023}.

\textbf{Yet given that AI proponents believe that the accelerated and ubiquitous adoption of AI above all else would demonstrate the US’s technological advantage and defense prowess over China, they position generalized benchmarks as a manufactured metric that serves to illustrate broad capabilities as an expression of such an advantage, a weak substitute for robust evaluation of what these systems can do when put into use.} These sentiments are further bolstered by drawing on parallels to the nuclear arms race against the Soviet Union, and applying the same logic to US adversaries by fueling concerns that China may outpace the US’s AI military capabilities through its investments in AI~\cite{work_beating_2019}. Yet the parallels between the nuclear arms race and AI have largely been found to be misrepresented, not only due to the lack of definable success of AI within military applications that have relied on promises and perception rather than capabilities, but because AI is not a weapon itself; it is, instead, a support or automation function. Indeed, unlike AI models, physical military systems like nuclear weapons can produce tangible illustrations of their capabilities. On the other hand, AI capabilities claims are difficult to substantiate not only due to their software-based nature~\cite{cummings_ai_2020}, but to a history of opaque and misleading claims that conceal the role of human involvement, making it difficult to ascertain and validate actual AI capabilities of state actors~\cite{Rollet_2025}.

Further challenges arise given that deep neural networks (DNNs), which form the basis of current foundation models, cannot solve tasks outside of their data distribution and training data sets. Their inability to handle novel scenarios that would arise from the fog of war thus raises serious questions about whether they can be successful in military settings. Appropriate evaluation metrics for AI’s fitness of purpose must consider how unreliable AI models may be in dynamic and uncertain environments, especially in situations not previously encountered, to reflect the true nature of warfare~\cite{cummings_ai_2020}. However, known deployments of military AI systems, including uses of foundation models by the Israel Defense Forces (IDF) in Gaza, have demonstrated a trend of fielding undertested AI systems in the battlefield for target recommendations, despite evidence of the constitutive lack of accuracy and known failure modes~\cite{biesecker_as_2025,dwoskin_israel_2024,abraham_lavender:_2024,gaza_algorithms_2024}.

Field testing is the process of evaluating the performance, reliability, and suitability of defense technologies under real-world conditions through war games, modeling, simulation, technical assessments, and, finally, live-force operations~\cite{noauthor_marine_nodate}. Once deployed, technologies are subjected to ongoing maintenance testing to ensure the continued reliability of their fitness for purpose. However, despite notable evidence of foundational models’ poor accuracy with non-English languages, especially for Arabic~\cite{naous_origin_2025,naous_having_2024}, forces such as the IDF continue to use foundation models to guide in the identification of targets, the most critical of defense operations, through means such as the faulty transcription and translation of intercepted communications in Gaza~\cite{biesecker_as_2025}. To make matters worse, the use of foundation models may feed into other AI-driven systems, as with the IDF’s use of Gospel, Lavender, and Where’s Daddy, which have contributed to the significant civilian death toll in Gaza through the fallible collection and use of intelligence information for targeting purposes, and which may even subject their users to violations of IHL~\cite{icrc_2024}. 

Indeed, the nested use of AI within decision-support systems will ultimately lead to compounding errors that propagate and accumulate through each subsequent phase of a kill chain~\cite{icrc_2024}, including for evaluating impacts of attacks required for field maintenance testing to ensure that technological outputs align with strategic and tactical defense objectives. Consider, for example, the IDF’s algorithm predicting how many civilians might be affected by attacks through the use of data from nearby cell towers to tally the number of civilians in an area~\cite{dwoskin_israel_2024}. This analysis in particular took “no account of whether a cellphone might be turned off or had run out of power or of children who wouldn’t have a cellphone.” Deploying AI models with profoundly low accuracy rates for target identification, compounded with flawed algorithms that are incapable of predicting the impact of AI systems’ decisions, not only impairs the ability to verify and validate AI-driven outcomes, but will result in consequences that may not be aligned with strategic and tactical defense objectives. This includes the potential loss of human control over the use of force in an armed conflict, and likely violations of IHL~\cite{viveros_alvarez_risks_2024}.

\textbf{Ultimately, the insistence on rapidly deploying foundation models in the highest-stakes defense applications (e.g., the DOD’s AICC), despite their demonstrable failure modes and clear lack of fitness for warfighting activities, illustrates how general benchmarking contributes to a manufactured narrative where the purported increase of broad (and unrelated) model capabilities have been falsely positioned as better performance indicators, antithetical to tried-and-true field-testing metrics.} Moreover, AI industry players further resist any form of ex ante scrutiny by positioning the kinds of evaluation and validation frameworks as overly burdensome, despite their wide utility across sectors~\cite{Malgieri_Pasquale_2022}. This has enabled systems to be deployed that perform at levels far below the risk thresholds that would be deemed acceptable through standardized safety processes in other domains.
\section{Looking Ahead: An Agenda to Correct Course}

The establishment of societally acceptable safety goals and tolerances to determine the risk thresholds against which technological systems can be assessed is an arduous task that can take years or decades to establish. Despite the nearly insurmountable challenges posed by such tasks, there are precedents to look to, such as the risk determinations established for nuclear systems against the backdrop of the Cold War, which have provided invaluable safety and dependability goals through which defense and safety-critical systems have been evaluated. This worthwhile effort has helped the US establish its technological advantage and defense prowess over its adversaries. But rather than follow such historical norms that necessitated broad democratic deliberations, AI technologists—primarily industry labs, as well as “AI safety”-focused organizations—have staunchly advocated for a skewed cost-benefit justification driven by the unsubstantiated claims of an AI arms race and speculative “existential” capabilities risks. Under such pretenses, specifically for the use of AI in defense and national-security applications, AI technologists have sought to substitute traditional safety frameworks with ill-defined “capabilities” or “alignment” counterparts that deviate from well-established military standards necessary for the functionality of defense systems. This “safety revisionism” may be precisely what disadvantages US military and technological capabilities against China or other adversaries by allowing for the accelerated AI adoption at the cost of lowered safety and security thresholds.

As evidenced by the slew of existing national-security concerns emanating from uses of AI models that are ripe for adversaries to exploit and compromise, ceding the determination of risk tolerance to AI technologists ultimately leads to the degradation of existing defense safety and security practices. Furthermore, this places decision-making regarding the acceptable societal risk tolerance for AI, including number of deaths and the corresponding safety methodologies (i.e., via dissemination of co-opted safety frameworks), in the hands of AI technologists themselves. As noted by Chauncey Starr, a pioneer of probabilistic risk analysis, such patterns are signifiers of an autocratic risk value system, a far cry from messaging that the acceleration of AI use within the US national-security apparatus would allow “democracies to maintain the lead”~\cite{amodei_trump_2025,murgia_anthropics_2024}. The utilization of said skewed risk thresholds within national security, in combination with the demonstrable brittleness of foundational models~\cite{zou_universal_2023,weidinger_ethical_2021}, further calls into question AI uses that may subject their users to violations of international humanitarian law~\cite{icrc_2024}, bring harms to both defense personnel and civilians in active war zones, and that ultimately normalizes fraught and inaccurate performance of AI in safety-critical contexts. Such normalization may imprint and set a precedent of low safety thresholds for civilian-critical infrastructure going forward (e.g., policing, energy, border governance, and so on). Indeed, as defense and civilian-safety critical standards are inextricably linked~\cite{NATO_joint_2018}, lowered safety thresholds for AI use within national security may imperil the safety of civil society given that military standards can set the precedent under which civilian safety-critical systems are to be evaluated.

Situating themselves as the arbiters of life or death, AI technologists have selected risk thresholds that utilize skewed risk and safety interpretations at odds with the demands of social, political, and ethical civilian life, and with established assurance practices critical to the safety and security posture of national and defense infrastructure. To remedy this gap, global governance efforts must establish consensus-building bodies that congregate civil society, legislatures, and judiciary bodies to establish democratically deliberated risk tolerances that require the consideration of the societal evaluation of risk and, as such, interpreting public attitudes and values. Governing bodies must move away from hollow “responsible” and “trustworthy” frameworks that concern broad and hypothetical capabilities and AI “alignment,” and maneuver toward defining AI risk thresholds that address “How safe is safe enough?” That is, they must produce AI safety objectives to safeguard defense and safety-critical systems in line with established norms, ultimately reckoning with the number of societally accepted fatalities and hazards that would require considerations such as to whom safety is afforded, under what contexts (e.g., human rights law versus IHL), and the exploration of concrete and substantiated AI risks.

Finally, policymakers and military stakeholders must preserve established military standards and risk thresholds, which have long ensured the safety and dependability of defense and critical systems, in spite of pressure to adopt manufactured metrics that serve to illustrate broad AI capabilities as an expression of military technological advantage. “Capabilities evaluations” and “red-teaming” are a weak substitute for existing TEVV frameworks that serve to evaluate a system’s fitness for purpose in line with strategic and tactical defense objectives. Indeed, with the emergence of increasingly powerful AI defense systems, from LAWS to AI DSS, we must ensure their deployment is not only aligned with risk thresholds that continue to maintain the safety and security of critical and defense infrastructure, but that they remain in line with IHL. The deadly and geopolitically consequential impacts of AI within military applications brings with it existential risks that are very real and present; as they stand, the safety interpretations put forward by AI technologists will only result in a significant civilian death toll with no meaningful oversight or validation regarding their use, reliability, or safety.

\bibliographystyle{plainnat}
\clearpage
\bibliography{citations}

\end{document}